\documentclass[prl,twocolumn,showpacs,preprintnumbers,amsmath,amssymb]{revtex4}
\usepackage{graphicx}
\usepackage{latexsym}
\usepackage{color}

\newcommand{\be}{\begin{equation}}
\newcommand{\ee}{\end{equation}}

 \newcommand{\bB}{\mathbf{B}}

\newcommand{\bj}{\mathbf{j}}

\newcommand{\br}{\mathbf{r}}
 \newcommand{\bv}{\mathbf{v}}

\begin{document}
\title{Dynamo action at low magnetic Prandtl numbers:\\
mean flow vs. fully turbulent motion}

\author{Y. Ponty$^1$ , P.D. Mininni$^2$, J.-F. Pinton$^3$, H. Politano$^1$ and A. Pouquet$^2$}
\affiliation{$^1$ CNRS UMR6202, Laboratoire Cassiop\'ee, Observatoire de la C\^ote d'Azur, 
BP 4229, Nice Cedex 04, France \\
$^2$NCAR, P.O. Box 3000, Boulder, Colorado 80307-3000, U.S.A. \\
$^3$CNRS UMR5672, Laboratoire de Physique, \'Ecole Normale 
Sup\'erieure de Lyon, 46 All\'ee d'Italie, 69007 Lyon, France}

\begin{abstract}
We compute numerically the threshold for dynamo action in Taylor-Green swirling flows. 
Kinematic calculations, for which the flow field is fixed to its time averaged profile,
are compared to dynamical runs for which both the Navier-Stokes and the induction equations are jointly solved.  The kinematic instability is found to have two branches, for all explored Reynolds numbers. The dynamical dynamo threshold follows these branches: at low Reynolds number it lies within the low branch while at high kinetic Reynolds number it is close to the high branch. 
%In the lower one, the dynamo threshold is equal to that of the dynamical problem in the laminar regime. The threshold for the higher branch is found to be  equal to that of the dynamical regime at large Reynolds numbers.
\end{abstract}
\maketitle

The magnetic field of planets and stars is believed to be the result of a dynamo instability 
originating in the flow motions inside their electrically conducting fluid core.  Dynamo occurs when induction due to motion overcomes diffusion~\cite{moffatt}, corresponding  to a threshold in the magnetic Reynolds number of the flow ($R_M=UL/\eta$, with $U$ and $L$  characteristic velocity and length scales of the flow, and $\eta$ the fluid's magnetic diffusivity).  For liquid metals (as molten iron in the Earth core, or liquid sodium in laboratory  experiments~\cite{GydroSpecialIssue}), the kinematic viscosity $\nu$ is several orders  of magnitude lower that the magnetic diffusivity $\eta$ -- the magnetic Prandtl number  $P_M = \nu/\eta$ is often of the order of $10^{-5}$ or lower.  As a result, the kinetic Reynolds number of dynamo generating flows tends to be very high.  Indeed, the relationship $R_V = UL/\nu = R_M/P_M$ implies that critical values $R_M^c$ of the order  of few tens are associated with Reynolds numbers in excess of one million.   This remark prompts two initial questions:  (i) can a dynamo instability develop from a fully turbulent flow?, and (ii) what is the evolution of the instability threshold as $R_V$ grows?  We partially addressed them in a previous numerical work~\cite{Paper1} for a flow generated by a deterministic Taylor-Green forcing at large scales.  Lowering the magnetic Prandtl number $P_M$ from $1$ to $10^{-2}$, we established that the laminar $R_M^c$ value undergoes an eightfold increase as unsteadiness and small scale  motion develop, and that once turbulence is fully established the threshold $R_M^c$ tends to saturate to a constant value. %In the process, the Reynolds number at onset varies from $30$ (at $P_M \sim 1$) to $13000$ (at $P_M \sim 0.01$).  
Dynamo action is thus preserved for flows having a well defined large scale geometry,  a question still open for turbulent motion without a mean flow~\cite{alex04}.   Numerical predictions of a dynamo threshold in realistic natural or experimental  conditions are still out of reach;  numerical simulations cannot get close to the parameter  range of planetary bodies or laboratory experiments (even if one ignores the question of boundary  conditions, a DNS study at $P_M \sim 10^{-5}$ would require a resolution in excess of $(10^5)^3$ grid points).  Nonetheless, the fluid dynamo experiments~\cite{KarlsruheRiga} in Riga and Karlsruhe found  the onset to be remarkably close to the values predicted from numerical simulations based  on the mean flow structure~\cite{GailitisStefaniTilgner}, and this despite the fact that the corresponding  flows are quite turbulent. It has led several experimental groups seeking dynamo action in less constraint geometry, eventually leading to richer dynamical regimes, to optimize the flow forcing using kinematic simulations based on mean flow  measurements~\cite{ForestKin,SaclayKin} -- with the advantage that mean flow profiles can be measured in the laboratory. It is thus of interest to test the validity of this procedure in numerical experiments.

In this Letter, we compare numerically the dynamo behavior as simulated from the dynamical magnetohydrodynamics (MHD) equations to the result of kinematic calculations in which the velocity is fixed to its time averaged profile.  In the fully dynamical problem, we integrate pseudospectrally the MHD equations : 
\begin{eqnarray}
&&\frac{\partial {\bf v}}{\partial t}+ {\bf v} \cdot \nabla {\bf v} =
-\nabla {\cal P} + \bj \times \bB +\nu \nabla^2 \bv + {\bf F} \label{E_MHDv} \\
&&\frac{\partial {\bf B}}{\partial t}+ {\bf v} \cdot \nabla {\bf B} = 
\bB \cdot \nabla {\bf v}
+\eta \nabla^2 {\bf B} \ , 
\label{E_MHDb}  
\end{eqnarray}
together with ${\bf \nabla} \cdot {\bf v} =0$, $\nabla \cdot {\bf B} =0$ 
and a constant mass density is assumed. Here, $\bv$ stands for the velocity field, $\bB$ the magnetic field, $\bj=(\nabla \times \bB)/\mu_0$ the current density  and ${\cal P}$ the pressure. In our case, the forcing term, ${\bf F}$, responsible for the generation of the flow is chosen to be the Taylor-Green vortex ($TG$)~\cite{meb}:
\begin{equation} 
{{\bf  F}_{\rm TG}(k_0)}= { 2F } \,  \left[ 
\begin{array}{c} 
\sin(k_0~x) \cos(k_0~y) \cos(k_0~z) \\ 
- \cos(k_0~x) \sin(k_0~y) \cos(k_0~z)\\ 0  
\end{array} \right]  \ , 
\label{eq:Ftg}
\end{equation} 
with $k_0=1$. For a given fluid viscosity and forcing amplitude, we first allow the flow to settle into a statistically steady state, in a nonmagnetic phase (Eq.~\ref{E_MHDv} with $\bB = {\mathbf 0}$). We either use Direct Numerical Simulations (DNS) for flows at $P_M$ of order one, or  Large Eddy Simulation (LES) schemes with an effective viscosity $\nu_{\rm eff}$ \cite{Ponty1} for $P_M$ lower than about 0.1. In the results reported here (cf. Table~I), the forcing flow amplitude is kept equal to 1.5, and the Reynolds number is increased by lowering the fluid's viscosity. Once the steady state is reached, a seed magnetic field with energy $10^{-20}$, evenly distributed among Fourier modes, is introduced and the MHD equations are integrated for several choices of the fluid's magnetic diffusivity. For each, we compute the exponential growth rate $\sigma_B = d(\ln E_B)/dt$.  The critical $R_{M,{\rm dyn}}^c$, for which $\sigma_B$ changes its sign, is thus obtained at constant $R_V$ in a  dynamical process for which the velocity field is a true Navier-Stokes flow. 
%The dynamo threshold is obtained with an uncertainty of the order of 10\%. 

In the kinematic study, we first compute the mean flow as an average in time of the velocity fields generated in the dynamical runs:
\begin{equation}
{\mathbf U}({\mathbf r}) = \frac{1}{T} \int dt  \, {\bv}({\mathbf r}, t) \ ,
\label{eqavg}
\end{equation}
where $T$ is a time scale chosen much larger than a typical eddy turnover time at the flow integral scale $L_{\rm dyn}$: $T_{\rm NL}= U_{\rm dyn}/L_{\rm dyn}$ ($U_{\rm dyn}$: characteristic velocity). Note that $T$ should exceed the magnetic diffusion time $T_{M} = R_M T_{\rm NL}$. In practice, an instantaneous velocity field  is extracted every time interval $\Delta T$ to increment the running average  of ${\mathbf U}({\mathbf r})$. We have used $\Delta T \sim T_{\rm NL}/100$, and $T > 200 T_{\rm NL}$.  To save computer time, the averaging is done during the linear growth (or decay) phase -- hence in the absence of a Lorentz feed-back in the Navier-Stokes equation. The induction equation~(2) is then solved  with $\bv({\mathbf r}, t)$ kept equal to ${\mathbf U}({\mathbf r})$, in search of growing solutions  ${\mathbf B}({\mathbf r}, t) = \exp \left( \sigma_B^{kin} t \right) {\mathbf B}({\mathbf r})$.  We note that the mean flow defined in Eq.~\ref{eqavg} is not a real flow in the sense that it is no longer a solution of the Navier-Stokes equations; for instance, it does not have a viscosity. We chose to attribute it the viscosity of the generating dynamical run. Then one could compute an associated kinematic Reynolds number $R_{v,kin} = U_{\rm kin} L_{\rm kin}/\nu$, but we have chosen to represent all Reynolds number variation as a function of $R_V \equiv R_{V,{\rm dyn}}$ (Figs. 1 and 3). Critical magnetic Reynolds numbers $R_{M,{\rm dyn}}^c$ and $R_{M,{\rm kin}}^c$ which are computed from each field's characteristic lengths (Fig. 2). 
%The threshold $R_{M,{\rm kin}}^c$ is again obtained as a marginal state between decaying and growing magnetic modes (here the uncertainty is of the order of a few percents). 

\begin{table}[htb]   
\centerline{
\begin{tabular}{|c|c|c|c|c|c|c|c|c|c|c|c|c|} \hline
$N$ & $\nu$ & $L_{\rm dyn}$ & $U_{\rm dyn}$ & $R_{V, {\rm dyn}}$ & $R_{M, {\rm dyn}}^c$ & $R_{M, {\rm kin}}^{c1}$ & $R_{M, {\rm kin}}^{c2}$ & $R_{M, {\rm kin}}^{c3}$\\ \hline
$64$ & $0.3$      & 3.08 & 1.44 &  14.81 & 34.39  & 21.95 & 53.16 & 142 \\ \hline 
%$64$ & $0.2$      & 3.29 & 1.82 &  30.14 & 28.82  & 25 & 50 & 150 \\ \hline 
$64$ & $0.1$     & 3.29 & 2.20 &   76.74 & 48.51  & 23.90 & 48.70 & 150 \\ \hline
$64$ & $0.08$    & 3.46 & 2.31 &  98  & 59.24  & 23.74 & 50.55 & 155 \\ \hline
$128$ & $0.04$    & 3.42 & 2.51 & 194   & 106.00 & 23.29 & 51.63 & 152 \\   \hline
$128$ & $0.015$   & 2.59 & 2.60 & 465   & 170.63 & 24.59 & 50.80 & 149 \\ \hline
$128$ & $0.01$   & 2.43 & 2.74 &  670   & 176.80 & 22.47 & 50.89 & 189 \\ \hline
$128$ & $8.6e^{-4}$  & 2.29 & 2.85 &  7954 & 150.42 & 23.48 & 51.24 & 167 \\ \hline
\end{tabular}
}
\caption{Parameters of the runs: grid resolution, fluid viscosity (effective value for the last  run which use LES), integral length and velocity scales, kinetic Reynolds number $R_{V,{\rm dyn}} = L_{\rm dyn} U_{\rm dyn} / \nu$, critical magnetic Reynolds number $R_{M, {\rm dyn}}^c$ for the dynamical problem and for the kinematic one: $R_{M, {\rm kin}}^{c1}$ for the onset of the the first dynamo mode, $R_{M, {\rm kin}}^{c2}$ the value at which it is no longer a growing solution and $R_{M, {\rm kin}}^{c2}$, the onset value for the second dynamo mode. The kinematic magnetic Reynolds numbers are computed using the kinematic integral velocity and length scales, and the magnetic diffusivity: $R_{M, {\rm kin}}^{ci} = L_{\rm kin} U_{\rm kin} / \eta^{ci}$. For all runs, we have measured $L_{\rm kin} \sim \pi$ and $U_{\rm kin} \sim 3.0$.}
\end{table}

\begin{figure}[t!]
\centerline{\includegraphics[width=8.5cm]{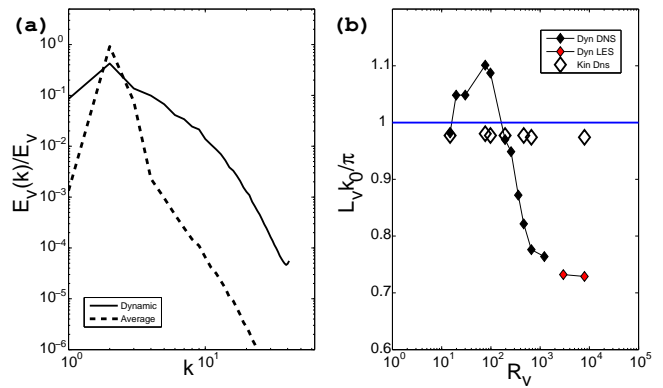}}
\caption{(a) kinetic energy spectra for TG1 ($\nu=0.01$); $E_{V, {\rm dyn}} (k,t=T)$ (solid line), $E_{V, {\rm kin}}(k)$ (dotted line). (b): integral length scales $L_{\rm dyn}$ and $L_{\rm kin}$, normalized by the size of the unit TG cell, versus the flow Reynolds numbers $R_V \equiv R_{V, {\rm dyn}}$. }
\label{spectra} 
\end{figure}

Before analyzing the dynamo behavior, we first compare characteristics of the dynamic and time-averaged velocity fields. Their spectra are shown in Fig.~\ref{spectra}(a), for fields originating in the DNS calculation at $R_{V, {\rm dyn}} = 670$. While the dynamical flow has a typical turbulence spectrum, the time-averaged field is sharply peaked at the size of the Taylor-Green cell. As the Reynolds number varies, the quantities for the average flow remain constant, while they do vary for the dynamical field. For instance, the flow integral scale is shown in Fig.~\ref{spectra}(b). It is computed from the kinetic energy spectra
\begin{equation}
\frac{L}{2\pi}= 
\left\langle \frac{\int dk \, E_{V}(k,t )/k}{\int dk \, E_{V}(k, t)} \right\rangle_T  
\label{eq:Ld}
\end{equation}
where $\langle \cdot \rangle_T$ stands for averaging in time. For Reynolds numbers less than about 100, $L_{\rm dyn}$ tends to be larger than the size ($\pi$) of one Taylor-Green vortex. At higher $R_V$'s, the turbulent flow has an integral length scale clearly confined within the Taylor-Green cell. On the opposite, the mean flow has $L_{\rm kin} \sim \pi$ and $U_{\rm kin} \sim 3.0$ at all $R_V$'s.

We now turn to the dynamo generation. In the fully dynamical problem, the $R_{\rm M, dyn}^c(R_V)$ curve -- Fig.2(b) -- displays an initial increase, corresponding to the development of turbulence, followed by a plateau~\cite{Paper1}. The dynamo threshold is then independent of the fluid's viscosity. This is understood if we recall that the dynamo is governed by large scale motions : indeed, in TG flows, we have reported~\cite{MininniAPJ} that in the early stage all Fourier modes grow at the same rate, but after a few hundred large scale turnover times only the modes near the TG forcing keep growing while the small scales are quenched. Besides, it is a common observation in fluid mechanics~\cite{PopeBook} that the geometry of a shear flow becomes  independent of the Reynolds number at high $R_V$, except for viscous sub-layers near  boundaries. One thus expects that, for flows having a well defined geometry, the dynamo persists at all (small) magnetic Prandtl numbers, with a finite value of $R_M^c$ as $R_V \rightarrow \infty$. See also~\cite{ShekoAPJ} and references therein for a recent discussion.  

\begin{figure}[t!]
\centerline{\includegraphics[width=6cm]{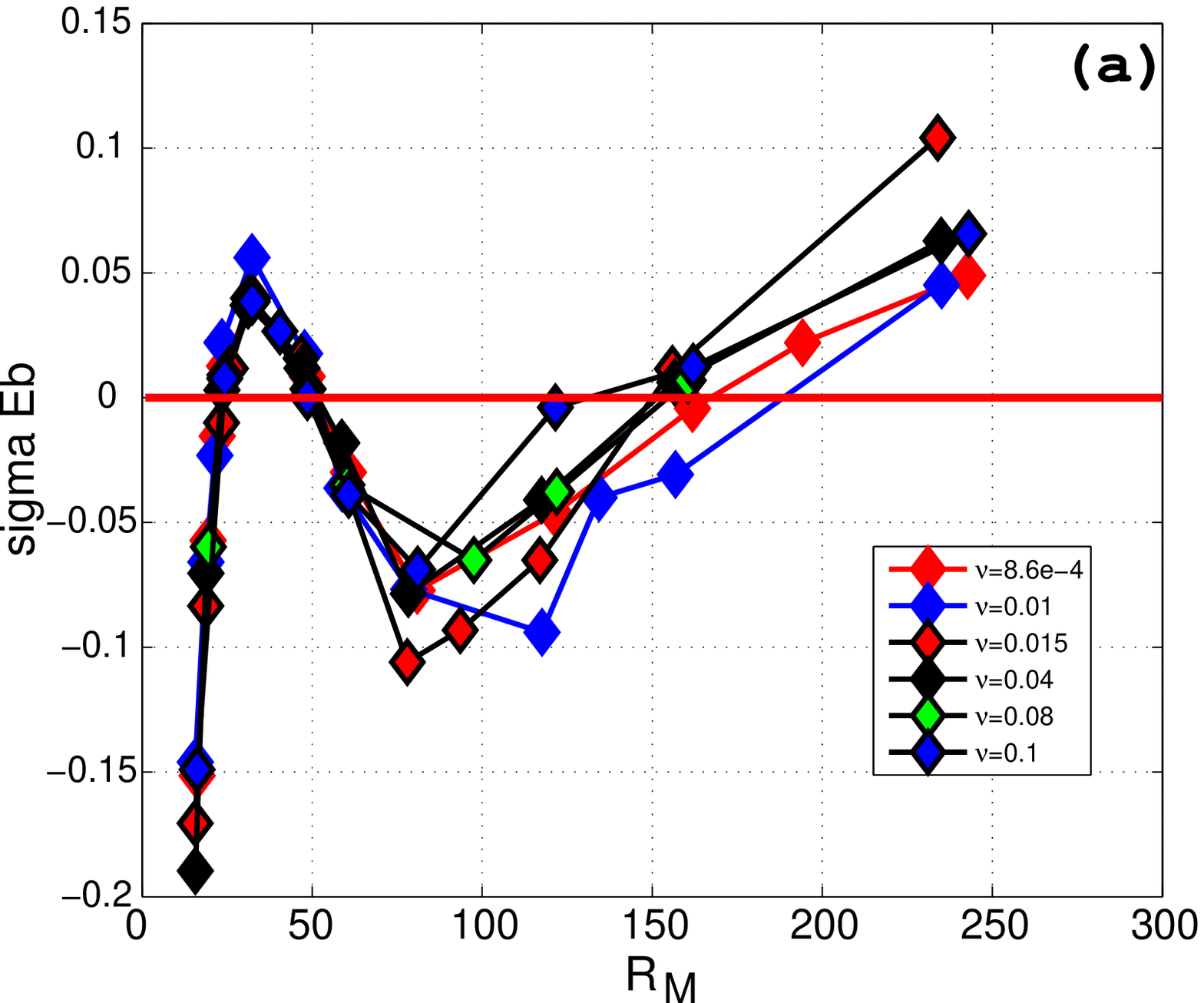}}
\centerline{\includegraphics[width=6cm]{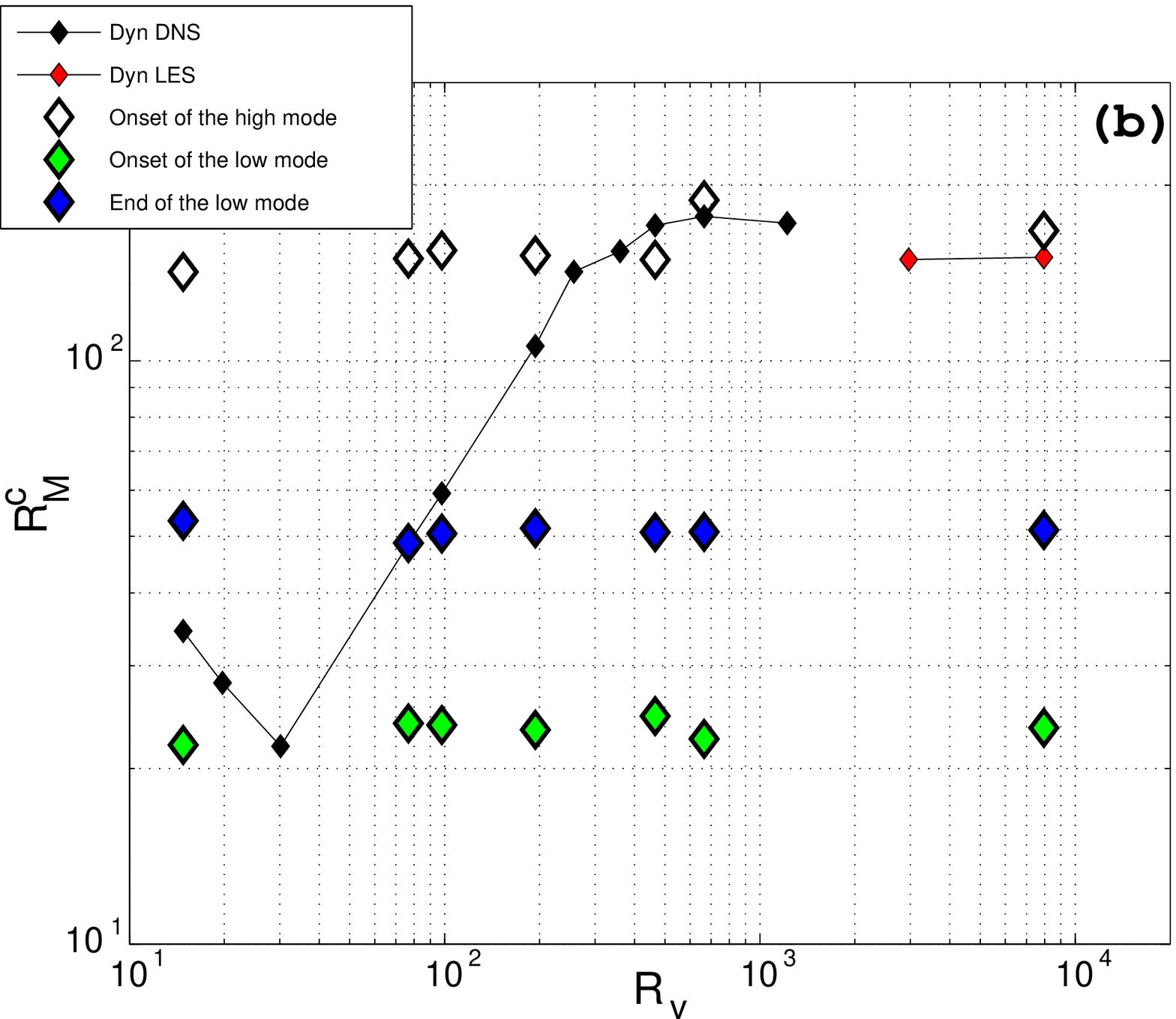}}
\caption{(a) Growth rates for the kinematic dynamo generated by mean flows computed for decreasing values of the fluid's viscosity. The intersections with the $\sigma_B = 0$ axis define the $R_{M, {\rm kin}}^{c 1,2, 3}$ values reported in Table~I. (b) Evolution of the critical magnetic Reynolds numbers $R_{M}^{c}$ with $R_V$. }
\label{growth} 
\end{figure}

For the time-averaged flow, we have observed the existence of two dynamo branches in the kinematic simulation of the induction equation -- a behavior already noted for the $ABC$ flow~\cite{GallowayFrisch}. As shown in Fig.~2(a), the kinematic growth rate is positive in the interval $[ R_{M, {\rm kin}}^{c1}, R_{M, {\rm kin}}^{c2} ] \sim [22, 51]$, and then again for  $R_{M, {\rm kin}} > R_{M, {\rm kin}}^{c3}$.  Beyond $R_{M, {\rm kin}}^{c3}$, the growth rate seems to be a monotonously increasing function of $R_M$. The first dynamo window is found to be essentially independent of the kinetic Reynolds number of the dynamic flow that has generated the time-averaged field. For the higher branch it varies within 15\% of a mean value $R_{M, {\rm kin}}^{c3} \sim 158$ (with no systematic trend in the range explored in our study). We compare in Fig.~2(b) the evolution of the critical magnetic Reynolds numbers. At low $R_V$, we observe that the dynamo threshold for the dynamical problem lies within the low $R_M$ dynamo window for the time-averaged flow. For $R_V$ larger than 200 the dynamical dynamo threshold lies in the immediate vicinity of  the upper dynamo branch (high $R_{M}^{c}$ mode of the time-averaged flow).

Let us now compare the structure of the resulting dynamo fields. The magnetic energies and corresponding integral length scales are shown in Fig.3. For the dynamical dynamo, energy is distributed in a broad range of scales. For the kinematic dynamos evolved from the time averaged flow, we observe that for the low $R_M$ mode the energy is strongly peaked at large scales, while it is more evenly distributed in the case of the high $R_M$ mode. This behavior is also reflected in the evolution of the integral magnetic scales, shown in Fig.~3(b). The kinematic low mode grows a dynamo essentially at scales {\it larger} than the Taylor-Green cell ($L_B \sim 1.6 \pi$) at all $R_V$'s. The high mode on the other hand grows within a TG vortex. The dynamo selected by the dynamical flow switches between these two behaviors. At low $R_V$ it grows with an integral scale larger than the TG cell and we observe that $L_{B, {\rm dyn}} \sim L_{B, {\rm kin}}$. At high $R_V$ the magnetic integral scale is about half the size of the TG cell. {The peak in the magnetic energy spectrum at smaller scales in the dynamic runs, as well as the smaller integral scale, suggest that turbulent fluctuations can play a role in the dynamo. This is in agreement with~\cite{MininniAPJ}, where both large and small scales were observed to cooperate in the dynamo process: the growth of small scales allows the dynamo to quench velocity fluctuations and the magnetic field to grow faster at large scales.}

\begin{figure}[htb]
\centerline{\includegraphics[width=8.5cm]{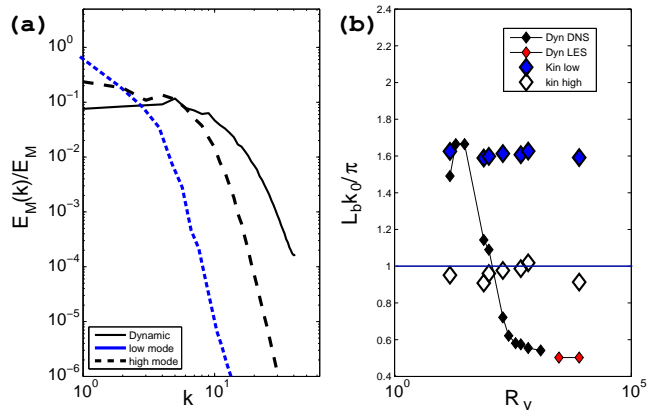}}
\caption{ (a): Magnetic energy spectra for $\nu=0.01$ (DNS at $R_V$=670);  (solid line) dynamical run, (short  and long dashed lines) low  / high $R_M$ kinematic dynamo modes (in this regime the spectra are computed as averages during the growth phase, normalized by the mean energy). (b) Evolution of the integral scales for the magnetic field, computed from the spectra.}
\end{figure}

The structure of the dynamos can be explored further with visualizations of the isosurfaces of the magnetic energy.  Fig.4(a) and (c) correspond respectively to the low and high  kinematic eigenfunctions, while Fig.4(b) and (d) show the dynamical fields, with the magnetic energy rescaled and averaged in time during the linear growth phase ({\it i.e.} $\langle E_M(\br, t) / E_M(t) \rangle_T$). Both a low ($R_V=76.74$) and a high ($R_V=670$) Reynolds number are shown. One observe a very good correspondence between the low $R_V$ dynamical mode and the kinematic low eigenfunction; indeed, at low Reynolds number the flow is laminar with small fluctuations about its mean. In this regime the dynamo is mainly generated in the shearing regions in between the TG cells~\cite{Nore,MininniAPJ}. At high $R_V$, we compare the dynamical growing dynamo to the structure of the high $R_M$ eigenfunction. In Fig.4(c) we recognize the  `twisted banana shape' structure of the neutral mode that underlies the $\alpha-\Omega$ dynamo of von K\'arm\'an flows (recall that at high $R_V$ the TG flow in each cell is very similar to the VK swirling flow)~\cite{Nore,SaclayKin}. For the dynamical flow at high $R_V$ the TG cells are no longer as coupled as they are at low $R_V$, and in Fig.4(d) one does not observe the clear pattern of the kinematic eigenmode. This can be due to the fact that the magnetic energy is displayed here during its initial growth. As seen in the spectrum-- Fig.3(a) -- the magnetic energy grows at all scales; it is only in the non-linear phase that the magnetic energy is eventually dominated by the large scales. However, at all $R_V$'s the dynamo grows predominantly in the planes $\pi/2$ and $3\pi/2$ which cut through the center of the Taylor-Green cells.  
 
\begin{figure}[htb]
\centerline{\includegraphics[width=8.5cm]{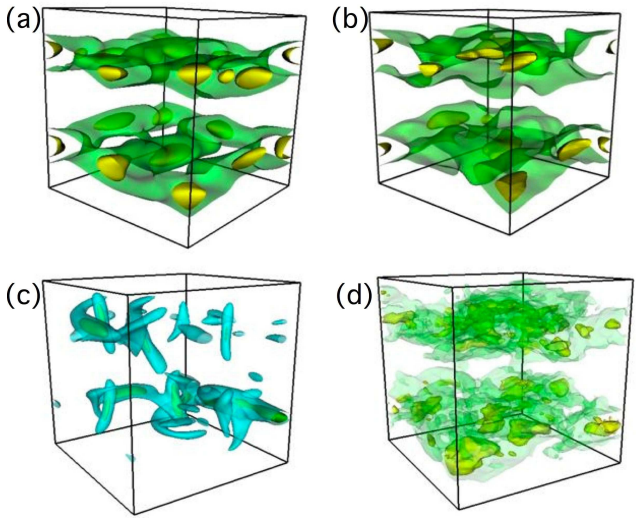}}
%\centerline{{\includegraphics[width=4.cm]{kin-Lowmode-iso0.5-iso0.75-2.eps}}
%{\includegraphics[width=4.cm]{Dyn-lowRv-iso0.5-iso0.75-2.eps}}}
%\centerline{{\includegraphics[width=4.cm]{kin-highmode-iso0.25-iso0.5.eps}}
%{\includegraphics[width=4.cm]{Dyn-highRv-iso0.5-iso0.75.eps}}}
\caption{Isosurfaces of the magnetic energy. (a): TG1 kinematic low eigenmode; (b): corresponding dynamical run at $R_V=76.74$; (c): high kinematic eigenmode; (d): corresponding dynamical run at $R_V=670$.  In each plot two isovalues have been plotted at 50\% and 75\% of the maximum magnetic energy -- except for the high kinematic modes in which the levels are 25\% and 50\%.} 
\label{isosurf}
\end{figure}

%The above observations (for instance the clear evolution of the dynamo integral length scale with $R_V$ and its link to the kinematic eigenmodes) may not hold for other types of flow forcing 
We have observed a similar behavior for flows forced with TG at $k_0=2$ (existence of at least two branches for the kinematic problem, with a transition of the dynamical dynamo threshold from the lower to the upper branch as turbulence becomes fully developed), but results may differ for other types of geometries or flow forcing~\cite{MinniMontgo}.  However we propose that the observations made here, particularly in the large $R_V$ limit, may have some relevance for laboratory experiments. For instance, we find that once turbulence is fully developed, the dynamo threshold is well approximated by the (high branch) kinematic value computed using the time averaged flow. This finding is in agreement with the observations in the Riga and Karlsruhe experiments, where the mean flow structure was optimized to favor dynamo action at low $R_M^c$. In addition the threshold reaches a limit value when $R_V$ increases, in agreement with kinematic simulations using von K\'arm\'an mean flows measured in the laboratory~\cite{SaclayKin}: the threshold was observed to be $R_V$ independent for $R_V$ exceeding about $10^5$. Hence one may expect that kinematic predictions based on hydrodynamic measurements  in laboratory prototypes can be relevant for experiments. This is also of interest for the numerical study of natural dynamos where a fully resolved description of the fluid's motion in the correct range of parameters is out of reach~\cite{Dormy}. There are however reasons to be cautious. The main concern lies in the observation that fully turbulent flows in confined volumes are not stationary. Long time dependence in the large scales velocity fluctuations  have been observed~\cite{VKS1,VolkFluct}. {Also, the effect of turbulent fluctuations on the threshold need further study. While recent works have show that a large scale incoherent noise may increase significantly the dynamo threshold~\cite{DubRecent}, small scale fluctuations can also be a source of dynamo action~\cite{moffatt,zeldo}.}

{\bf Acknowledgements. \ }%\acknowledgments
We thank F. Daviaud, B. Dubrulle, R. Volk and A. Schekochihin for fruitful discussions. NSF ARI grant CDA--9601817 and NSF grant CMG--0327888 (NCAR) are acknowledged. JFP, HP and YP thank CNRS Dynamo GdR, and INSU/PNST and PCMI Programs. Computer time was provided by NCAR, PSC, NERSC, and IDRIS (CNRS). 

%%%%%%%%%%%%%%%%%%%%%%%%%%%%%%%%%%%%%%

\end{document}